# Dynamic Labyrinthine Pattern in an Active Liquid Film

Yong-Jun Chen[1], Yuko Nagamine[2] and Kenichi Yoshikawa[1,2*]

[1]Department of Physics, Graduate School of Science, Kyoto University, Oiwake-cho, Kitashirakawa, Sakyo-ku, Kyoto 606-8502, Japan

[2]Spatio-Temporal Order Project, ICORP, JST (Japan Science and Technology Agency), Kyoto 606-8502, Japan

*Corresponding author: yoshikaw@scphys.kyoto-u.ac.jp

## Abstract

We report the generation of a dynamic labyrinthine pattern in an active alcohol film. A dynamic labyrinthine pattern is formed along the contact line of air/pentanol/aqueous three phases. The contact line shows a clear time-dependent change with regard to both perimeter and area of a domain. An autocorrelation analysis of time-development of the dynamics of the perimeter and area revealed a strong geometric correlation between neighboring patterns. The pattern showed autoregressive behavior. The behavior of the dynamic pattern is strikingly different from those of stationary labyrinthine patterns. The essential aspects of the observed dynamic pattern are reproduced by a diffusion-controlled geometric model.

PACS number: 47. 15. G-, 47. 20. Dr, 47. 54. –r, 47. 55. np

## Introduction

Pattern formation is ubiquitous in nature. Labyrinthine pattern formation is found in a variety of chemo-physical systems, such as reaction-diffusion systems, a thin magnetic film, an amphiphilic Langmuir monolayer, type I superconductors in a magnetic field, a droplet of ferrofluid in a magnetic field and Hele-Shaw flow [1-3]. Generally, labyrinthine pattern formation arises due to competition between two antagonistic interactions [2], which have been interpreted theoretically in the framework of reaction-diffusion equations [4]. The system tries to achieve a minimum energy and stationary pattern corresponds to the state at the minimum energy [3]. The labyrinthine pattern formed in the aforementioned systems maintains its morphology as the pattern evolves. The final pattern depends on the initial perturbation. However, to our best knowledge, a persistently dynamic labyrinthine pattern in a liquid film is not reported. In this paper, we report spontaneous dynamic labyrinthine patterns in inertial liquid-liquid dewetting.

When a liquid film is deposited on a nonwettable substrate, the liquid will dewet and leave a dry patch. Dewetting has been studied intensively because of its technological importance [5]. A hole forms and grows, and the liquid nucleates as dewetting proceeds. For a thin film (submicrometer thickness), viscosity and intermolecular interaction dominate the dewetting process [6]. When a film has a relatively large thickness $h$ (millimeter-scale), an inertial effect is present not only in wetting [7] but also dewetting [8, 9]. Gravity creates static pressure and tends to flatten the liquid on the substrate. The driving force for dewetting is a negative spreading coefficient [8]. If the thickness $h$ is below a critical thickness $h_c$, a liquid film with a relatively large thickness (millimeter-scale) will spontaneously dewet when a hole larger than some critical size is formed in the film [8, 9]. In an inactive system of inertial liquid-liquid dewetting, a hole maintains its shape during growth on a uniform substrate [5]. A capillary wave and gravity wave have been found to be characteristics of the inertial effect. The front of the hole grows constantly while obeying Culick law for the bursting of a soap film [9]. After the film reaches its critical thickness $h_c$, the dewetting process stops. The liquid film takes partial wetting on the substrate. However, in an active system as that of the present article, the system agitates in a spontaneous manner [10-11]. Chemical nonequilibricity causes persistently dynamic motion of contact line. Here, we describe complex dynamic labyrinthine patterns formed through an active dewetting process in an alcohol/water system.

## Experiment

We used pentanol (Wako, Japan, density: $\rho_o = 0.810 g/ml$, organic: o) and pure water (w) in the experiment. Pentanol is partially miscible with water (Concentration $c \le 2.7 vol\%$ at $20\,°C$). A circular glass Petri dish was filled to a depth of several centimeters with pure water and this was covered by a thin film of pentanol (millimeter thickness). The thickness of pentanol film is less than the critical thickness for dewetting. We found that the critical thickness $h_c$ of a pentanol film on pure water (contaminated by pentanol) is about 1.60 mm. A hole larger than a critical size was created by blowing on the air-organic interface using a pipette. To control the thickness of the pentanol film, we withdrew pentanol from the film using a glass syringe. We performed experiments using different sizes of Petri dishes with diameters in the range of 10-20 cm. The system was illuminated by an optical-fiber lamp and the evolution of the contact line was projected onto tracing paper. The map of the contact line on the tracing paper was monitored by a video camera. The experimental setup and a typical pattern are shown in Fig. 1. The phenomenon was analyzed using image-analysis software.

## Results

Figure 2 exemplifies the growth of a hole when a hole is created in a thin film with a thickness smaller than the critical thickness for dewetting. In the early stage (Inset in Fig. 2), the radius of the hole grows at a constant velocity of 2.14 cm/s, which obeys Culick law [9]. However, after the hole reaches its maximum radius, the contact line does not stop, but rather returns toward the air-water interface. The hole shrinks and instability of the contact line grows. The contact line then develops complex patterns, as shown in Fig. 2. Based on a measurement of geometric parameters (perimeter and area of the hole) (Fig. 2), we found that the hole exhibits oscillating behavior. The behavior of the contact line is markedly different from that of a water film dewetting from the hydrophobic liquid perfluorodecalin [9]. After the hole grows, the dynamic labyrinthine pattern exists for a long time (on the order of hours). The morphology of pattern changes continuously (movie 2 [12]). Figure 3 shows typical dynamic patterns formed by the contact line. The contact line forms an interesting labyrinth, where the labyrinthine pattern changes dynamically with time. The dynamics of the patterns is rather different from those of stationary patterns in systems mentioned previously [2]. Figure 4 shows the dynamics of a single-domain pattern. The perimeter of the pattern changes with time. As the pattern evolves, it moves within the Petri dish. The trajectory of the center of mass is shown in Fig. 4(a). In the experiment, we found that the pattern tends to move by backing on the wall of the Petri dish. The velocity of the center of mass is on the order of centimeters per second. The perimeter and area of the domain are plotted in Fig. 4(b). The dynamics of the patterns are continuous. The evolution of the pattern exhibits periodic behavior because of interaction with the wall of the Petri dish. The system is far from equilibrium. The contact line moves continuously. We found that contact lines repulse each other. When two contact lines approach each other, damping interaction exists between them. Two contact lines do not fuse into each other unless they move toward each other at large velocity. The distance between two nearest stable contact lines is about 1 mm. This is similar to the case of interacting chemical fronts in the chemical reaction-diffusion system described by Lee *et al* [1]. Thus, a contact line can evolve into a complex labyrinthine structure. The pattern can be replicated by splitting of a domain, and two domains can coalesce. Such coalescence of contact lines creates new domains and islands within a domain.

## Discussion

As a pattern evolves, the contact line does not maintain its morphology, unlike the formation of a steady pattern [1-3]. Patterns with various morphologies were found, such as long fingered, labyrinthine, and branching structures. As shown in Fig. 4(b), perimeter and area of the pattern change constantly. During pattern evolution, pattern dynamics shows periodic behavior, as in Fig. 4(b). We examine the time-correlation of patterns. The autocorrelation function of patterns can be written as

$$C(T) = \frac{< [G(t) - \overline{G}][G(t+T) - \overline{G}] >}{\sigma^2}$$

where <> represents a mean value, $G$ represents a geometric parameter of a pattern (perimeter or area), $\overline{G}$ is the mean value of the perimeter or area, $\sigma^2 = < (G(t) - \overline{G})^2 >$ is the variance of the geometric parameters, $t$ is time and $T$ is the time lag. The autocorrelation functions of the perimeter and area in Fig. 4(b) are plotted in Fig. 4(c). The autocorrelation shows that neighboring patterns are stongly correlated to each other. Pattern evolution is predictable and not random. A negative correlation suggests oscillation of the perimeter and area [13]. The autocorrelation function reflects the periodic evolution of patterns in Fig. 4(b). This is consistent with the Fourier transformation of the time-traces of the perimeter and area in Fig. 4(d). Several peeks are noted in Fig. 4(d), indicating the frequencies of oscillation. This oscillation of the perimeter and area reflects the periodic evolution of the energy of the system. The periodic shape of the autocorrelation function in Fig. 4(c) suggests that the patterns show autoregressive behavior of patterns evolution [13]. This periodic time-evolution of geometric parameters should be related to the interaction with the no-flux wall, the inertial effect and the change in the surface tension of the air-water interface. As shown in Fig. 2, when a hole reaches its maximum radius due to dewetting, the direction of the driving force, which governs the motion of the contact line, changes. This change in the direction of the driving force leads to a dewetting-wetting transition.

The initial thickness of the film is important in dynamic pattern formation.

The effective tension in a film is defined as $\gamma_{eff} = \gamma_{a/o} + \gamma_{o/w} - \gamma_g$, the value $\gamma_g = \frac{1}{2}\tilde{\rho} g h^2$ is the tension due to the static pressure of gravity [5], where $\tilde{\rho} = \rho_o(1 - \rho_o / \rho_w)$, $\rho_w$, $g$, and $h$ are the effective density, density of water, acceleration due to gravity and thickness of the film, respectively, and the subscripts a, o, and w represent air, organic phase (pentanol) and aqueous phase, respectively. The effective tension $\gamma_{eff} = \gamma_{a/o} + \gamma_{o/w} - \gamma_g$ tends to induce dewetting while the surface tension of the air-water interface $\gamma_{a/w}$ tends to make the film thin (Fig. 1). The critical initial thickness for the dewetting of pentanol from water can be determined by the equation $\gamma_{eff} = \gamma_{a/w}$. Therefore, the critical initial thickness for dewetting is

$$h_c = \sqrt{(\gamma_{a/o} + \gamma_{o/w} - \gamma_{a/w}) / (1/2\tilde{\rho} g)} = \sqrt{-S/(1/2\tilde{\rho} g)}$$

where the spreading coefficient $S = \gamma_{a/w} - \gamma_{o/w} - \gamma_{a/o}$. Our experiment showed that the critical initial thickness for the growth of a hole is about 1.60 mm. Thus, we obtain a spreading coefficient of $S = -1.92 mN/m$ with the above relationship. However, if the velocity of dewetting at an early stage in Fig. 2 agrees with Culick law $V = \sqrt{|S|(1 - h^2/h_c^2)/2K\rho_o h}$ (inertial effect is not included) ($V = 2.14 cm/s$), where $V$ is velocity and $K = 2.0$ because the velocity is small and within the quasistatic regime [9], we can estimate the critical initial thickness $h_c' = 2.10 mm$ and thus obtain a spreading coefficient of $S' = -3.31 mN/m$ (using the above expression for critical thickness). The pseudo-error of the estimation based on Culick law indicates that the inertial effect contributes to the spreading coefficient. The contribution of the inertial effect arises from the static pressure of gravity $\Delta S = S - S' = 1/2\tilde{\rho} g(h_c^2 - h^2) = 1.50 mN/m$ ($h = 1.40 mm$ in Fig. 2). We estimate the interfacial tension at the organic-water interface $\gamma_{o/w}$ according to Ref. [14] using the ratio of the adhesion and cohesion of pentanol. A theoretical estimation gives $4.12 mN/m$ for the surface tension at the pentanol-water interface $\gamma_{o/w}$ (for pure water, $\gamma_{a/w} = 72.80mN/m$ and $\gamma_{a/o} = 25.80mN/m$ at 20°C [14]). Consequently, the surface tension in the air-water interface (contaminated) that is freshly exposed to air is 28.00 mN/m with $S = \gamma_{a/w} - \gamma_{o/w} - \gamma_{a/o}$. Upon exposure to air, the surface tension in the air-water interface is determined by diffusion, dissolution and volatilization. The surface tension of air-water interface depends on the area of the air-water interface that is exposed to air. The area of the air-water interface depends on the total volume of pentanol, since the dewetting process tends to minimize the free energy of the system. In the experiment, we found that a pattern with a very large domain tends to split. Thus, multiple domains coexist when the initial film is thin. Islands of pentanol are found in a domain during evolution of the domain when the initial film is thin. The size of the dish also affects the pattern morphology. In a small Petri dish, only one domain usually forms when the initial thickness is large, while multiple domains coexist in a large Petri dish. The pattern from a thicker initial film is usually simpler than that from a thinner initial film.

To observe the actual behavior of pentanol on pure water, we place a droplet of pentanol on pure water. The droplet spreads completely and disappears. Apparently, pentanol completely wets the surface of pure water. However, when we place a droplet of pentanol on the surface of aqueous solution with 2.3 *vol%* pentanol, the droplet does not disappear soon, but moves spontaneously as reported previously [10]. The surface tension of the air-water interface is decreased by the dissolving of pentanol in pure water. In the experiment, we found a broad class of substances that exihbit similar behavior, including nitrobenzene, aniline, chlorobenzene and pentanol [11, 15]. Pseudo partial wetting of these substances on an air-water interface was found [11]. A monomolecular layer spreads from a droplet on an aqueous solution to the air-water interface. The molecules dissolve and volatilize from the air-water interface. This spreading process changes the surface tension of the air-water interface. Marangoni-driven spreading induces the self-propelled motion of a droplet on an aqueous solution [11]. Since the water surface has a large surface tension, it is surprising that a pentanol film dewets from the surface of pure water (see Fig. 2). The driving force for wetting and dewetting is the spreading coefficient [5]. For pentanol on pure water, the spreading coefficient $S = \gamma_{a/w} - \gamma_{a/o} - \gamma_{o/w} = 40mN/m > 0$ ($\gamma_{a/w} = 72.8mN/m$ for pure water). The dewetting of pentanol from the water surface indicates that the surface tension of the water surface, which is exposed to air after dewetting, is less than that of pure water as mentioned earlier. The surface tension of the contaminated water surface is close to that of a saturated aqueous solution. However, when the area of the hole is large enough, the surface tension will recover to that of pure water through the dissolution and vaporization of molecules. We observed a dewetting-wetting transition of film under the assistance of an inertial effect.

The morphology of the patterns formed in our system is similar to that of the viscous fingering patterns in a Hele-Shaw cell [16]. Figure 5 shows the evolution of fingers in a pentanol film. The growth of fingering patterns is similar to that in Saffman-Taylor instability [16]. A small instability evolves into fingers. However, in the last stage of the fingering growth in Fig. 5, the fingers are smoothed out due to the flow in the pentanol film. The fingers cannot grow continuously because there is no external pressure and the patterns are confined to the film. The flow and pressure in the film due to gravity couple the motion of different parts of the contact line. The nature of this coupling is unclear. Our system is an open space without external pressure, while a Hele-Shaw cell is a closed space. The thickness of our pentanol film is not constant. A gradient of thickness in the film will induce fluid flow in the pentanol film. Thus, the mechanism of our dynamic pattern formation is different from that of the Saffman-Taylor fingering pattern. The competition between the effective tension in the film and the surface tension in the air-water interface controls the morphology of the pattern. The continuous dissolution and volatilization at the air-water interface cause the dynamic evolution of patterns. The motion of the contact line depends on the concentration of pentanol in the vicinity of the contact line. The chemical nonequilibricity near the contact line at the air-water interface disturbs the dynamic motion of the contact line. The dynamic phenomenon observed in the experiment is due to chemical nonequilibricity in the air-water interface.

To obtain insight into the physics of dynamic pattern formation, we consider the fluid field in the pentanol film and the concentration field of pentanol in air-water interface. In the pentanol film, we use Darcy law to describe the motion of the fluid (Reynolds number is less than ten). The velocity field of a fluid in a liquid film is proportional to the gradient of pressure in the fluid [16], i.e, Darcy law: $\vec{v} = -\frac{e^2}{12\mu}\nabla P$, where $\mu$ is the fluid viscosity in an organic film. With the continuity condition $\frac{\partial h}{\partial t} + \nabla \cdot \rho\vec{v} = 0$, we obtain $\frac{\partial h}{\partial t} = \frac{h^2}{12\mu}\rho\nabla^2 P$, where $P = P_0 + \frac{1}{2}\rho g h^2 - \gamma_{a/o} C_s$, $C_s$ is the curvature of the air-organic interface and $P_0$ is the pressure in air. Thus, we obtain an equation for the evolution of the thickness of the film:

$$\frac{\partial h}{\partial t} = \frac{h^3}{12\mu}\rho g \nabla^2 h + \frac{h^2}{12\mu}\rho g \nabla h \cdot \nabla h - \frac{\rho h^2}{12\mu}\gamma_{a/o}\nabla^2 C_s \qquad (1)$$

On the other hand, the density of alcohol molecules in the air-water interface $\Gamma$ can be characterized by a diffusion equation [17]:

$$\frac{\partial \Gamma}{\partial t} = D\nabla^2 \Gamma - j(t,x,y) \qquad (2)$$

where $D$ and $j(t,x,y)$ are a diffusion coefficient and the net molecular flow of vaporization and dissolution from the air-water interface, respectively. The evolution of the two phases of the pentanol film and the air-water interface is coupled through the contact line. It is a challenge to solve this free boundary problem, not only analytically but also numerically. However, Eq. (1) and Eq. (2) suggest that the motion of the boundary between the pentanol film and the air-water interface is determined by the diffusion process in the two areas. The driving force for motion of the contact line is related to how pentanol diffuses away and how pentanol fluid flows in and out at the vicinity of the contact line. We can make an analogy to the front motion of solidification in crystal growth [18]. In solidification, the temperature field is governed by the diffusion of heat [18]. In the vicinity of the front, diffusion of heat is related to the curvature of the interface between the solid and liquid. Pattern formation in solidification can be described using a boundary-layer model with diffusion-controlled evolution [19, 20].

If we consider the evolution of a single domain pattern for simplicity, the equation that governs the motion of the contact line is

$$\frac{d}{dt}\frac{\partial L}{\partial \dot{\vec{r}}(s)} - \frac{\partial L}{\partial \vec{r}(s)} = -\frac{\partial \Re}{\partial \dot{\vec{r}}(s)} \tag{3}$$

and the Lagrangian $L$ can be expressed as $L = \int_0^{s_0} ds[\frac{1}{2}|\dot{\vec{r}}|^2 \rho_l - E(\vec{r}(s))]$, where we parameterize the closed boundary of the pattern as $s \in [0, s_0]$, with generalized coordinates $\vec{r}(s)$ and velocity $\dot{\vec{r}}(s)$, and $s$, $s_0 = s_0(t)$, $\rho_l$ and $E(\vec{r}(s))$ are the arc length of the contact line, the total arc length, the density of contact line and the potential experienced by the contact line, respectively. In the Lagrangian $L$, we include the inertial effect of motion of the contact line. The energy dissipation due to viscosity is expressed by the Rayleigh function [21] $\Re = \frac{\zeta}{2}\int_0^{s_0} ds|\dot{\vec{r}}|^2$, where $\zeta$ is a friction coefficient. From Eq. (3), we obtain

$$\rho_l \frac{dU}{dt} = -\zeta U - \hat{n} \cdot \nabla E \tag{4}$$

where $U = \dot{\vec{r}} \cdot \hat{n}$ and $\hat{n}$ is a normal unit vector on the contact line. The force $-\hat{n} \cdot \nabla E$ is determined by the concentration field in the air-water interface and the thickness field in the film. The force that acts on the contact line is orthogonal to the tangential vector of the contact line. The general kinetic dynamics of a curve can be described using equation [22]

$$\dot{\kappa} = -[\kappa^2 + \frac{\partial^2}{\partial s^2}]U \tag{5}$$

where $\kappa$ is the curvature of the contact line. The perimeter and area of contact line are not conserved. To solve Eqs. (4) and (5), we must specify $-\hat{n} \cdot \nabla E$.

The surface concentration of pentanol in the air-water interface is controlled by a spreading process. The spreading rate of pentanol at the air-water interface is determined by the diffusion of pentanol at the air-water interface. As shown by a previous calculation [10], diffusion depends on the shape of boundary. The gradient of concentration is proportional to the curvature of contact line [10]. This is similar to the diffusion-controlled solidification and melting in crystal growth [18, 20]. The motion of the interface is driven by its curvature. Thus, we can write the driving force due to curvature-governed diffusion acting on the contact line as [18]

$$-\hat{n} \cdot \nabla E = \alpha\kappa + \Lambda_b(A_0 - A) \tag{6}$$

where $\alpha$ is an adjustable constant. Interaction with the no-flux boundary of the Petri dish conserves the volume of the pentanol film. Global feedback adjusts curvature-driven evolution and causes periodic oscillation of the pattern. The effect of the no-flux boundary and the effect of gravity, which depends on the thickness of the film, can be modeled as $\Lambda_b(A_0 - A(t))$ in Eq. (6), where $\Lambda_b$, $A_0$, and $A$ are a coupling constant, equilibrium area and instantaneous area of the domain. For a closed curve, the dynamics of the area can be written as $\frac{\partial A}{\partial t} = \oint_l dsU$, where $l$ is a closed curve [22]. In addition, a non-equilibrium spreading process disturbs the motion of the contact line. The external disturbance is one of the reasons for the dynamic behavior of the pattern. For simplicity, we only consider the inertial effect in our calculation.

By substituting Eq. 6 into Eq. 4, we have (by setting $\alpha = 1.0$)

$$\kappa = \rho_l \dot{U} + \zeta U - \Lambda_b(A_0 - A) \tag{7}$$

Using Eq. (5), we obtain

$$\rho_l \ddot{U} + \zeta \dot{U} + \Lambda_b \oint_l dsU = -[\rho_l \dot{U} + \zeta U - \Lambda_b(A_0 - A)]^2 U - \frac{\partial^2 U}{\partial s^2} \tag{8}$$

In addition, we set $U(s)\big|_{t=0} = 0$ and the initial condition of the contact line

$$x(\theta)\big|_{t=0} = \cos\theta[R_0 + \varepsilon\cos(f\theta)]$$

$$y(\theta)\big|_{t=0} = \sin\theta[R_0 + \varepsilon\cos(f\theta)]$$

where $x$ and $y$ are orthogonal coordinates as shown in Fig. 6 and $\theta \in [0, 2\pi]$, $R_0 = 2.0$, $\varepsilon = 0.1$, $f = 9.0$ are the angle in polar coordinates, the initial radius, the intensity of the disturbance and the frequency of the disturbance of the circle. Figure 6 shows the numerical results of Eq. (7) and Eq. (8). With a small initial disturbance, curvature drives the evolution of the contact line. An inertial effect that arises from motion of the contact line causes pattern dynamics. The morphology of the domain changes during evolution. In the numerical simulation, we consider the coalescence of contact lines and repulsive interaction between two contact lines (they are not modeled in the mathematic model). The pattern will split during evolution, as shown in Fig. 6, and the small part is omitted in the result. Thus, there is a shift in the area and perimeter when a domain splits. Figure 7 shows the evolution of the curvature of the boundary. The Long- and short-term evolution of curvature clearly indicate dynamic behavior of the pattern [Fig. 7(a) & 7(b)]. Comparing with the evolution of a front of solidification, the inertial effect causes a dynamic evolution of morphology [18]. Figure 8(a) shows the perimeter and area of the domain. Periodic behavior is noted due to interaction from the no-flux wall and conservation of the volume of the pentanol film. In the actual motion of the contact line, the effect of no-flux boundary of the Petri dish is complex and not uniform along the contact line. Interaction with the no-flux boundary induces oscillation of the pattern near a certain equilibrium position. However, we found that the coupling constant $\Lambda_b$ is the dominant factor for the period of oscillation in the numerical simulation, as shown in Fig. 8(b). The oscillation of area strongly depends on the coupling constant $\Lambda_b$, but not linearly. The perimeter of domain boundary also shows certain periodic behavior depending on the coupling interaction. The numerical results qualitatively reproduce the dynamic behavior of the pattern. However, the curvature-driven growth of the pattern is complex. The specification of force $-\hat{n}\cdot\nabla E$ as a function of curvature in Eq. 4 has an important effect on the morphology of the pattern [22]. Disturbance from nonequilibricity is not included in our simulation. As shown previously, the nature of disturbance determines the dynamic behavior of pattern evolution in a neural system [23]. It would be interesting to include more information in the model and this would be worthy of further study.

## Conclusion

In summary, we have demonstrated a dynamic labyrinthine pattern in an active liquid system of pentanol/water. The patterns show oscillating behavior and autoregressive behavior. The motion of contact line is controlled by diffusion process. The pentanol/water system is an active system that demonstrates self-agitation. The experiment showed that self-agitation caused by the chemical nonequilibricity of surface tension in the vicinity of the contact line plays an important role in dynamic behavior. A curvature-governed theoretical model has been proposed. The effects from the boundary of Petri dish and the inertial effect of motion are considered, which cause dynamic evolution of the morphology of patterns. Our theoretical model reproduces the essential aspects of dynamic patterns. However, dissolution and volatilization at the air-water interface as continuous disturbances change the surface density of molecules dynamically and have important effect on the dynamic behavior of patterns. The effect of external disturbance on the pattern evolution is the future project. The importance of external disturbance in the dynamic pattern formation has been noted in a theoretical model of neural population [23]. It is the center of interest to understand dynamic pattern generation in behavior and neural systems [24-25]. Our results may contribute to the understanding of dynamic patterns in biological systems.

Acknowledgement

We are grateful to Professor Chad Leidy (University of the Andes, Colombia) for his helpful discussion and Dr. Sumino for a critical reading of the manuscript. Useful advice from Dr. Ma Yue is also gratefully acknowledged. Y. J. C. thanks Ichikawa foundation for fellowship. Y. J. C. also is supported by Global COE program (The Next Generation of Physics, Spun from University and Emergence) of Graduate School of Science, Kyoto University. This work was supported by a Grant-in-aid for Creative Scientific Research (Project No. 18GS0421).

Figure captions

FIG. 1 (Color online) Schematic of the experimental setup. A Petri dish is filled with a layer of pure water, which is covered by a thin liquid film of pentanol before (up) and after (down) a hole is created in the pentanol film. At the right side, a typical pattern is shown. Scale bar is 2cm. The inset is drawing of contact area of air, pentanol and water. The surface tensions ($\gamma_{a/o}, \gamma_{o/w}$ and $\gamma_{a/w}$) and the effective tension from gravity ($\gamma_g$) are indicated by arrows.

FIG. 2 (Color online) Growth of a hole. The upper images are a series of snapshots of the growth of the hole. The lower panels are plots of the area and perimeter of the hole against time. The inset is plots of the area and perimeter at an early stage of hole growth. The lines are fitted to the data. The initial thickness of the film is 1.40 mm. The diameter of the Petri dish is 18.0 cm. Scale bar is 3cm. For details, see Movie 1 [10].

FIG. 3 Dynamic labyrinthine patterns. After the initial growth of a hole, a dynamic labyrinthine pattern emerges. In this figure, we show a typical example. The initial thickness of the film is about 1.50 mm. The diameter of the Petri dish is 18.0 cm. Scale bar is 3cm.

FIG. 4 (Color online) Dynamics of a single domain pattern. The initial thickness of the film is 1.50 mm. The diameter of the Petri dish is about 11.0 cm. (a) A pattern and superimposed trajectory of the center of mass of the patterns. (b) Plot of the perimeter and area of the patterns against time. (c) Autocorrelation function C(T) of the perimeter (P) and area (A). (d) Fourier transformation of time traces of the perimeter and area in (b) (FTP for perimeter and FTA for area). For details, see Movie 3 [10].

FIG. 5 Snapshots of the evolution of fingering patterns. The initial thickness of the film is 1.50 mm. The diameter of the Petri dish is 11.0 cm. Scale bar is 2cm. For detail, see Movie 4 [10].

FIG. 6 (Color online) Numerical results for the time-development of a single domain pattern. We set $\alpha = 1.0$, $\rho_l = 1.0$, $\zeta = 0.5$, $\Delta t = 0.002$, $\Lambda_b = 0.2$ and $A_0 = 20.0$ in the simulation. For detail, see Movie 5 [10]. Splitting of a domain and the coalescence of contact lines are indicated by circles and arrows.

FIG. 7 (Color online) Time-evolution of the curvature of the contact line. (a) Long-term evolution of curvature. (b) Short-term evolution of curvature. We set $\alpha = 1.0$, $\rho_l = 1.0$, $\zeta = 0.5$, $\Delta t = 0.002$, $\Lambda_b = 0.2$ and $A_0 = 20.0$ in the simulation. For detail, see Movie 6 [10].

FIG. 8 (Color online) Perimeter of the contact line and area of a single domain pattern. (a) Oscillation of the perimeter and area ($\Lambda_b = 0.2$); (b) Dependence of the oscillation of the perimeter and area on the coupling constant $\Lambda_b$. The corresponding results for $\Lambda_b = 0.01$, $\Lambda_b = 0.05$, $\Lambda_b = 0.10$, $\Lambda_b = 0.15$ are indicated by arrows. We set $\alpha = 1.0$, $\rho = 1.0$, $\zeta = 0.5$, $\Delta t = 0.002$ and $A_0 = 20.0$ in the simulation.

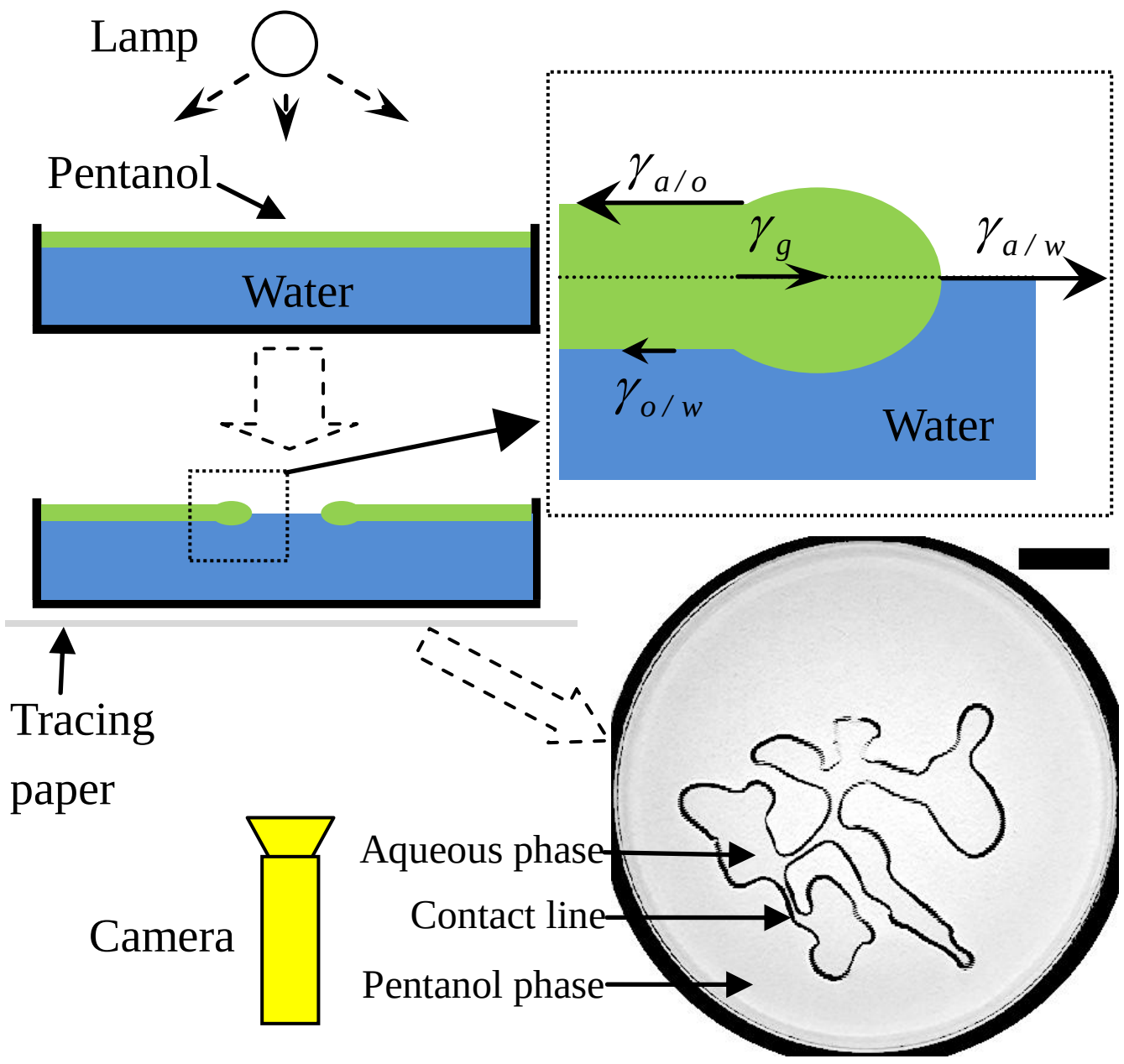

Figure 1 Yongjun Chen, Yuko Nagamine, Kenichi Yoshikawa

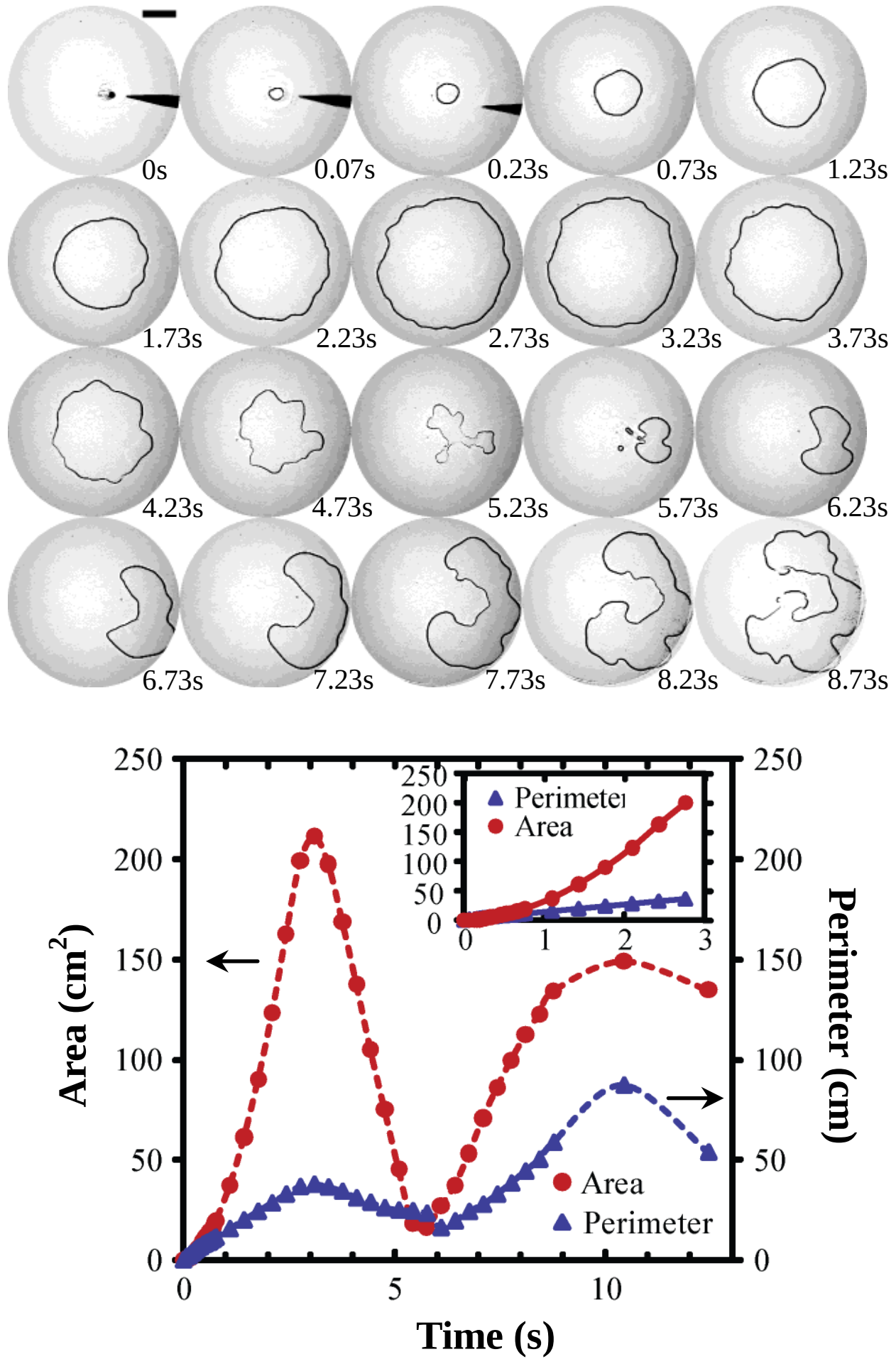

Figure 2 Yongjun Chen, Yuko Nagamine, Kenichi Yoshikawa

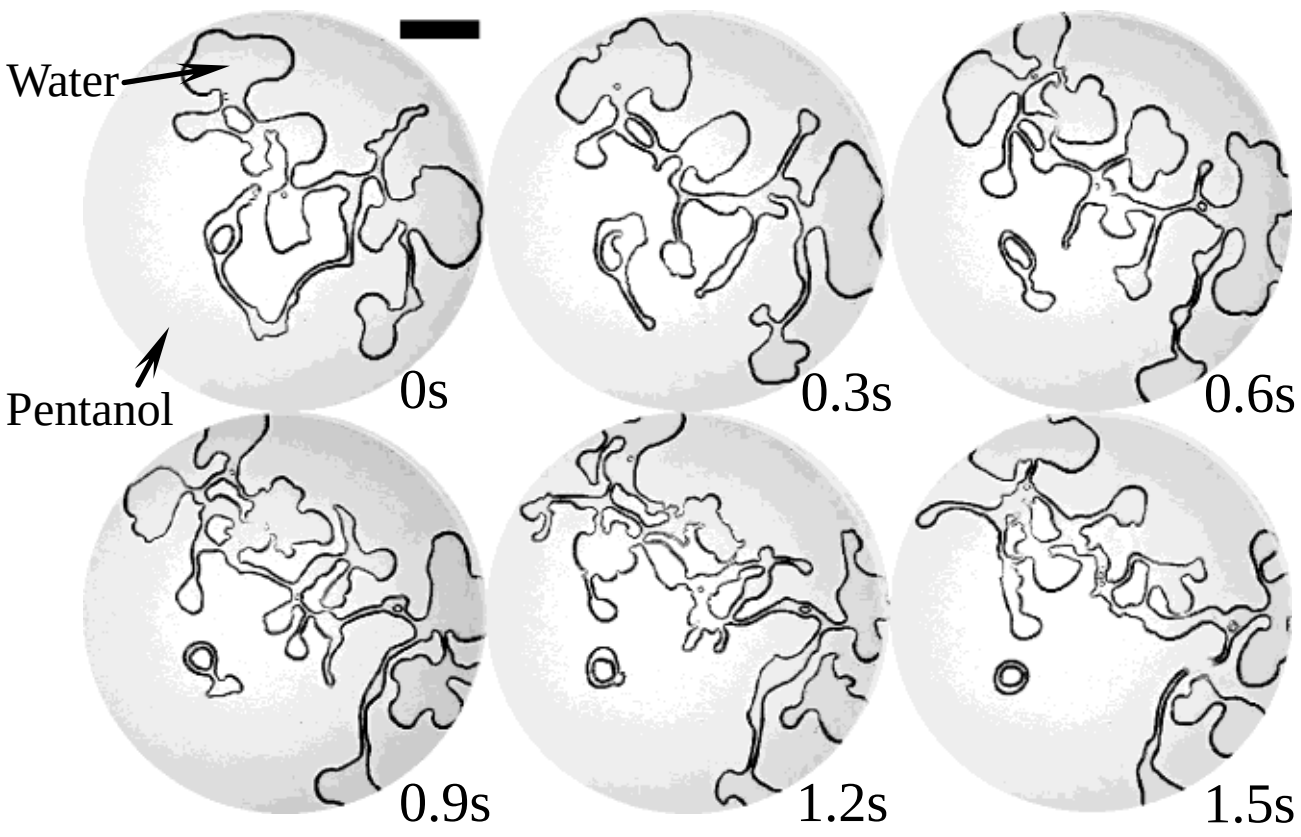

Figure 3 Yongjun Chen, Yuko Nagamine, Kenichi Yoshikawa

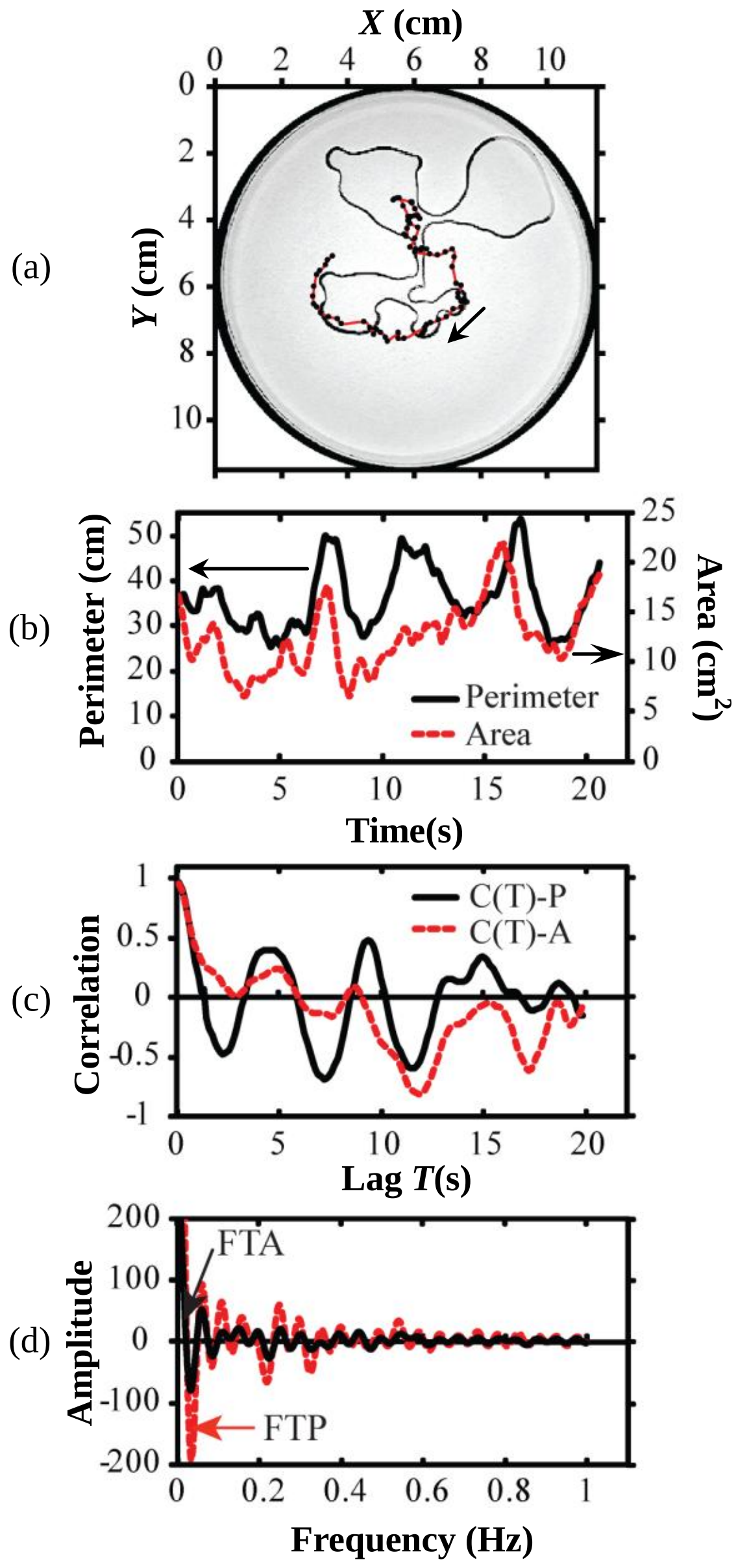

Figure 4 Yongjun Chen, Yuko Nagamine, Kenichi Yoshikawa

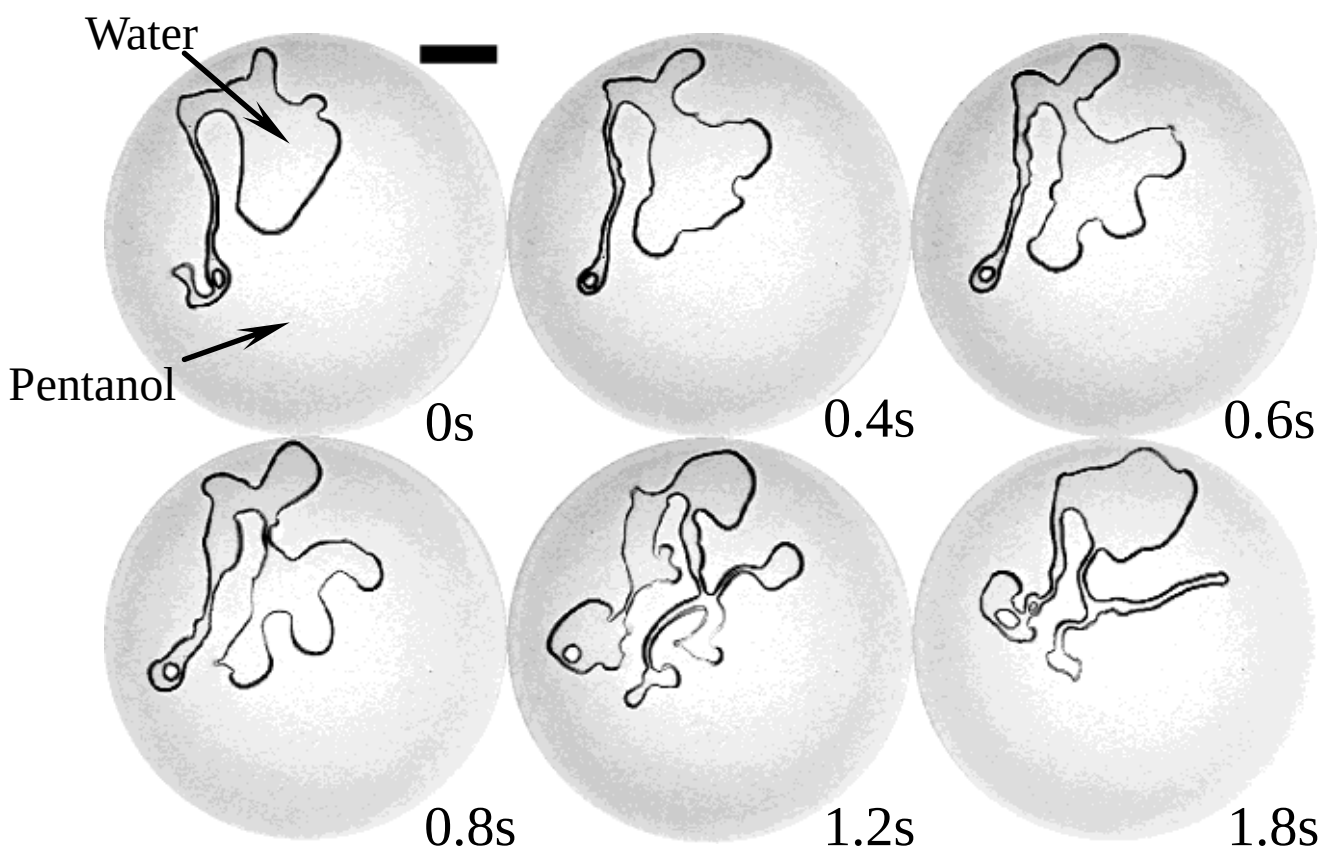

Figure 5 Yongjun Chen, Yuko Nagamine, Kenichi Yoshikawa

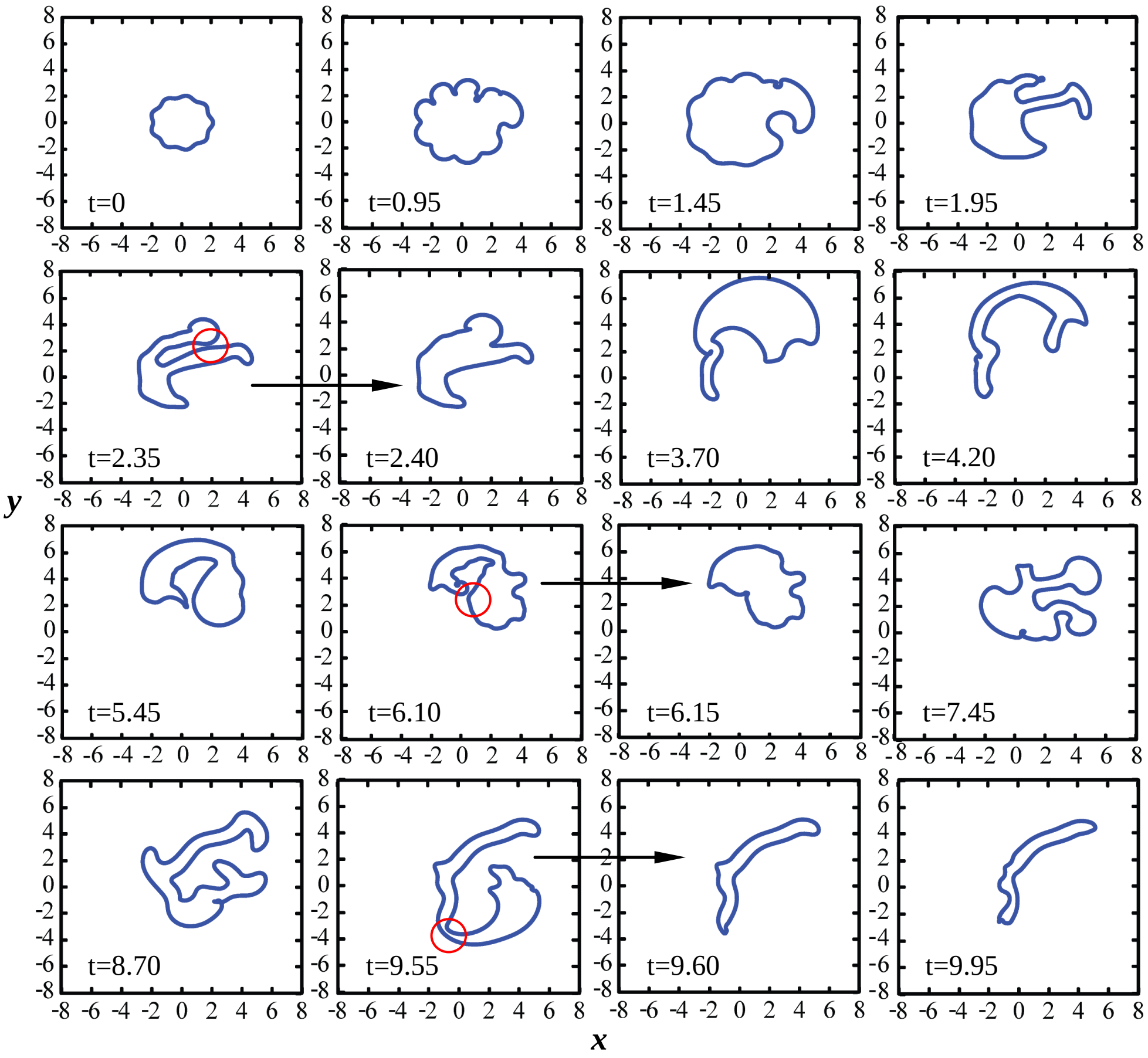

Figure 6 Yongjun Chen, Yuko Nagamine, Kenichi Yoshikawa

(a)

(b)

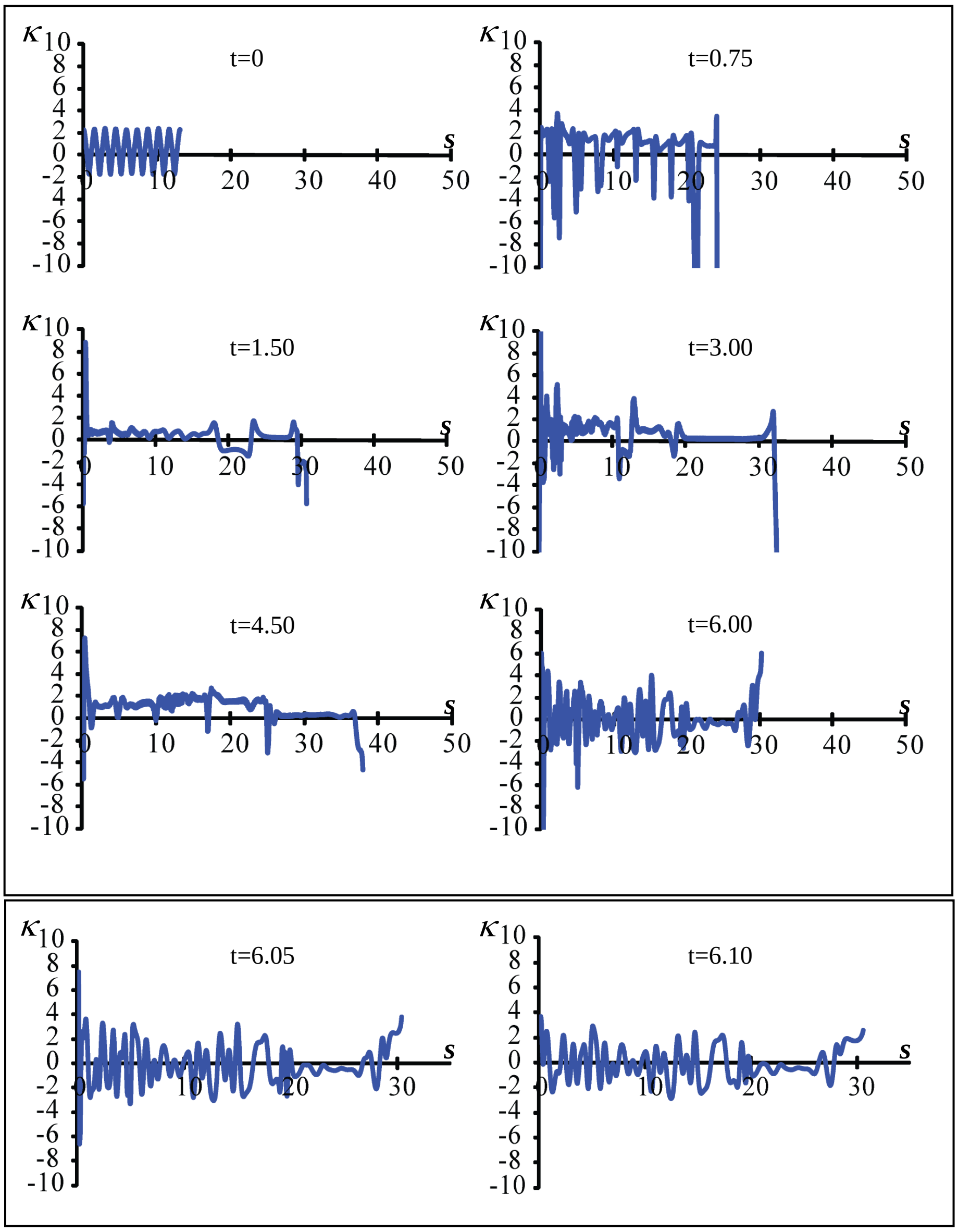

Figure 7 Yongjun Chen, Yuko Nagamine, Kenichi Yoshikawa

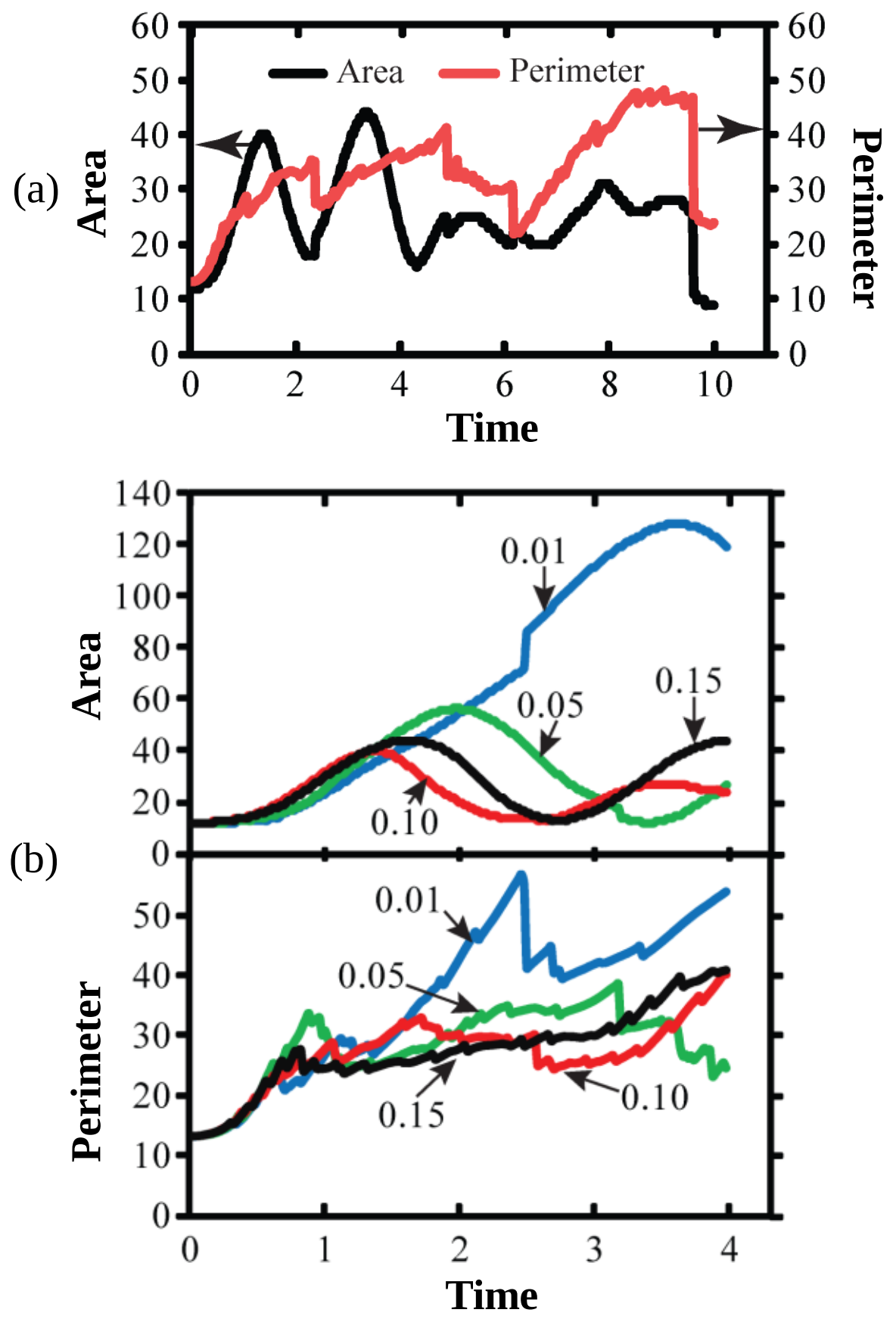

Figure 8 Yongjun Chen, Yuko Nagamine, Kenichi Yoshikawa